\documentclass[aps,prb,twocolumn,amsmath,amssymb,superscriptaddress,showpacs]{revtex4}

\usepackage{amsmath}
\usepackage{graphicx}

\begin{document}

\title{Trends in bulk electron-structural features of early transition-metal carbides}  
\author{Aleksandra Vojvodic}
\email{alevoj@chalmers.se}
\author{Carlo Ruberto}
\affiliation{Department of Applied Physics, Chalmers University of
  Technology, SE-412 96 G\"oteborg, Sweden}

\date{\today}


\begin{abstract}

A detailed and systematic density-functional theory (DFT) study of a
series of early transition-metal carbides (TMC's) in the NaCl
structure is presented. The focus is on the trends in the electronic
structure and nature of bonding, which are essential for the
understanding of the reactivity of TMC's. The employed approach is
based on a thorough complementary analysis of the electron density
differences, the density of states (DOS), the band structure, and the 
real-space wave functions to gain insight into the bonding of this
class of materials and get a more detailed picture of it than
previously achieved, as the trend study allows for a systematic
identification of the bond character along the different bands. Our
approach confirms the presence of both the well-known TM--C and TM--TM
bonds and, more importantly, it shows the existence and significance
of direct C--C bonds in all investigated TMC's, which are frequently
neglected but have been recently identified in some cases 
[Solid State Commun. \textbf{121}, 411 (2002);  
Phys. Rev. B \textbf{75}, 235438 (2007)]. New information on the
spatial extent of the bonds, their \textit{k}-space location within
the band structure, and their importance for the bulk cohesion is
provided. Trends in covalency and ionicity are presented. The
resulting electron-structural trends are analyzed and discussed within
a two-level model.  

\end{abstract}


\maketitle


\section{Introduction}

An extensive study of the reactivity of the transition-metal carbide
and nitride (TMX) surfaces is being performed with the
density-functional theory 
(DFT).\cite{RuVoLu06,VoRuLu06,RuVoLu07,RuLu07,VoHeRuLu,VoRuLu} It is
aimed at an understanding of reactivity from fundamental principles,
similar to the one available today for pure metal
surfaces.\cite{BlNobook08} Due to the intimate relation between bulk
and surface electronic structures, a careful mapping of trends in the
bulk electronic structure is essential for the overall purpose to 
understand the trends in reactivity of the transition-metal carbides
(TMC's). Therefore, a bulk background focusing on the trends in
electronic structure and nature of bonding in the early
transition-metal carbides (TMC) is here provided.   

The relevance of the results is, however, broader, as the bulk TMX's
are also interesting by themselves, having such properties as
extremely high melting point, ultrahardness, and metallic
conductivity.\cite{Oy96,Pi96} As a consequence, the importance of the
electronic-structure trends for atomic structure and stability is also
analyzed. 

The bulk TMX's have been studied experimentally and theoretically with
a large number of
techniques.\cite{NeRaEiWeSch76,GeWiMo83,Sch87,ZhGuJeChAn88,PrCo89,GuGr89,HaGrJaGu91,GuHaGr92,GuHaGr93,HaGuGrKo93,Jo95,GrMiCoCoLo99,ZhLiZhXi02,DaDeMo05,ViSoLiRoIl05,RuLu07} 
Already the early studies agree on the fact that the bonding in these
compounds involves simultaneous contributions from metallic, covalent,
and ionic bonding.\cite{Sch87} The main electronic-structure
properties are related to 
(i) the direction and amount of charge transfer between the TM and X
atoms, responsible for the ionicity of the material, and  
(ii) the modifying effect on the metal \textit{d} band upon the
carbide/nitride formation, responsible for the formation of hybridized
bonding and antibonding $pd$ states. The decreasing stability of the
TMC's in NaCl structure from left to right along a period in the
periodic table has been explained as arising from the successive
filling of the antibonding TM$d$--C$p$ states in a rigid-band 
model.\cite{HaGuGrKo93,GeWiMo83,PrCo89} 

Recently, the contributions to the cohesive energy from the TM--C,
TM--TM, and C--C bonds, respectively, have been approximately
determined quantitatively by a two-sublattice model of the NaCl structure
based on DFT calculations.\cite{ZhLiZhXi02} These results confirm the
dominance of the TM--C bond (strongest for TiC) and the importance of
the TM--TM bond (strongest for VC). In addition, they show that direct
C--C interactions cannot be neglected in some of the considered TMC's
(from CrC to NiC along period 4 in the periodic table). The used
model, however, falls short of explaining such results from the
calculated electronic structures. 

\begin{figure}
\includegraphics[width=.25\textwidth]{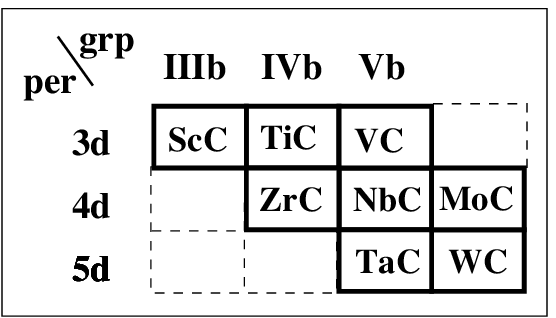}
\caption{\label{fig:TMCs}
The early transition-metal carbides under investigation, organized
according to their parent-metal position in the periodic table.}
\end{figure}

Our systematic DFT investigation is performed on a number of
non-magnetic early TMC's. Complementary electronic-structure analysis
tools are employed in a joined way to obtain detailed information of
the bond character: valence-electron density differences, Bader charge
analyses, band structures, projected densities of states (DOS), and 
real-space wave functions. Key bulk properties are calculated,
compared with earlier studies, and analyzed on the basis of the
obtained knowledge of the electronic structure. The studied subgroup
of TMC's (Fig.~\ref{fig:TMCs}) is chosen to allow a monitoring of
trends with respect to the constituent transition metals (TM) along
both period and group within the periodic table. Also, some of these
TMC's are widely used in applications. We argue that the subgroup is
large enough to pick up the important trends.   

Many TMC's crystallize in the sodium chloride structure, either as
stable, including ScC (at normal pressures), TiC, VC, ZrC, NbC, and
TaC, or metastable compounds, such as $\delta$-MoC and WC. The choice
of the TMC's in the NaCl structure is natural for our calculations
since the number of varying parameters should be kept small when
performing a trend study to allow for an identification of the
important electronic factors.     

The Paper is organized as follows. First, the computational details
are summarized in Sec.~\ref{sec:comp_det}. Then, in
Sec.~\ref{sec:res}, the results from our calculations are presented
and the obtained trends analyzed, including: atomic geometry and
stability (Sec.~\ref{sec:bulk_geometry}), electron density and charge
transfer (Sec.~\ref{sec:dens_diff_Bader}) and detailed electronic
structure (Sec.~\ref{sec:DOS_bulk}). In Sec.~\ref{sec:disc}, these
trends and the nature of bonding in the TMC's are summarized,
discussed, and related. The main conclusions of our study are
summarized in Sec.~\ref{sec:conc}.  

%
%



\section{Computational details}\label{sec:comp_det}

The calculations presented in this paper are performed within the
well-established DFT formalism using the plane-wave pseudopotential
code Dacapo.~\cite{dacapo} The ion-electron interaction is treated
with Vanderbilt ultrasoft pseudopotentials.~\cite{Va90} The
exchange-correlation energy is included by the generalized gradient
approximation (GGA), using the PW91 functional.~\cite{BuPeWa98} All
the bulk calculations are performed with a Monkhorst-Pack
sampling~\cite{MoPa76} of $8\times8\times8$ special \textit{k}-points
and a cutoff energy for the plane-wave expansion of 400~eV. The
similarities and differences in the nature of bonding in the
considered TMC's are studied with the following electron-structure
analysis tools: electronic density differences, Bader charge
analyses,\cite{Ba91,HeArJo06} band structure diagrams, atom-projected
local DOS's, and Kohn-Sham wave functions. The Bader analysis is a
charge-localization analysis that is able to give quantitative
information of the electron localization around different ions. The
scheme used here is based on the approach described in
Refs.~\onlinecite{Ba91} and~\onlinecite{HeArJo06}.


\section{Results and Analysis}\label{sec:res}

\subsection{Geometry and Stability}\label{sec:bulk_geometry}

\begin{figure*}
\centering
\includegraphics[width=\textwidth]{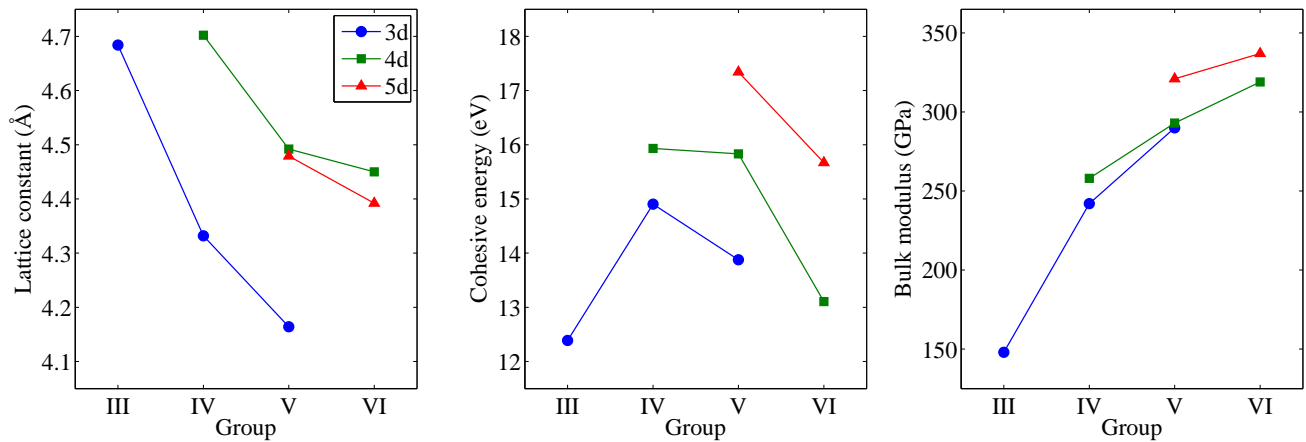}
\caption{\label{fig:bulk}
  Variations of, from left to right, the lattice parameters, the
  cohesive energies, and the bulk moduli of the investigated $3d$,
  $4d$, and $5d$ TMC's as a function of the group number of their
  parent transition metals.}
\end{figure*}

In Table~\ref{tab:TMC_a0} and Fig.~\ref{fig:bulk} we present
the calculated lattice parameters, bulk moduli, and cohesive energies,
together with the available experimental data.
The lattice parameters and the bulk moduli are obtained with a
Murnaghan equation of state~\cite{Mu44} and the cohesive energies are
calculated as 
\begin{equation}\label{eq:E_cohesive}
E_{\text{coh}}=E_{\text{solid}}-\sum_\text{atoms}
E_{\text{atom}}^{\text{isolated}}.
\end{equation}
%

%
\begin{table}
\centering
\caption{\label{tab:TMC_a0}
  Calculated lattice parameters ($a_0$), cohesive energies
  ($E_{\text{coh}}$), bulk moduli ($B_0$), and Bader charge transfers
  from TM to C atoms (Bader). The corresponding experimental values
  are given in parentheses. The calculated $E_{\text{coh}}$ values are
  defined by Eq.~(\ref{eq:E_cohesive}).} 
\begin{tabular}{ccccc}
\hline
\hline
TMC & $a_0$ & $E_{\text{coh}}$& $B_0$ & Bader  \\ 
    & (\AA) & (eV/unit cell)& (GPa) & ($|e|$/atom)  \\
\hline
ScC & 4.684 (4.637)\footnotemark[1] & 12.39 (12.74)\footnotemark[4] & 148 (--)  & 1.54   \\
TiC & 4.332 (4.325)\footnotemark[2] & 14.90 (14.32)\footnotemark[4] & 242 (232-390)\footnotemark[3] & 1.49  \\
VC  & 4.164 (4.163)\footnotemark[2] & 13.88 (13.88)\footnotemark[4] & 290 (308-390)\footnotemark[3] & 1.41  \\
\hline
ZrC & 4.702 (4.691)\footnotemark[2] & 15.93 (15.86)\footnotemark[5] & 258 (159-224)\footnotemark[3] & 1.70   \\
NbC & 4.492 (4.454)\footnotemark[2] & 15.83 (16.52)\footnotemark[5] & 293 (296-330)\footnotemark[3] & 1.64  \\
$\delta$-MoC & 4.450 (4.270)\footnotemark[5] & 13.11 (14.45)\footnotemark[5] & 319 (--)  & 1.97  \\
\hline
TaC & 4.479 (4.453)\footnotemark[2] & 17.34 (17.12)\footnotemark[6] & 321 (214-414)\footnotemark[3] & 1.94  \\
WC  & 4.382   (--)  & 15.67 (16.49)\footnotemark[6] & 357 (--)  & 1.60  \\
\hline
\hline
\end{tabular}
\footnotetext[1]{Ref.~\onlinecite{DaDeMo05}.}
\footnotetext[2]{Ref.~\onlinecite{NaYa05}.}
\footnotetext[3]{Refs.~\onlinecite{Pi96} and \onlinecite{ViSoLiRoIl05}.} 
\footnotetext[4]{Ref.~\onlinecite{HaGrJaGu91}.}
\footnotetext[5]{Ref.~\onlinecite{GuHaGr92}.}
\footnotetext[6]{Ref.~\onlinecite{GuHaGr93}.}
\end{table}
%

%
%

There is good agreement between calculated and experimental
values. For the lattice parameters the deviations are smaller than
$1\%$. Our calculated lattice constants are also in quantitative
agreement with other published theoretical results, obtained with both
the PW91 and the RPBE GGA functionals.\cite{ViSoLiRoIl05}

For both the cohesive energies and the bulk moduli the deviations from
experiments are larger. Still, our results agree well with other
theoretical values.\cite{GrMiCoCoLo99,ViSoLiRoIl05,DaDeMo05} The
deviations from experiments can be understood from the fact that in
reality the TMC's often occur as substoichiometric phases, with
strongly varying amounts of carbon vacancies. It is known that several
bulk properties depend strongly on such defects.\cite{Jo95,DrBoRuAo95}

\begin{figure}
\centering
\includegraphics[width=.50\textwidth]{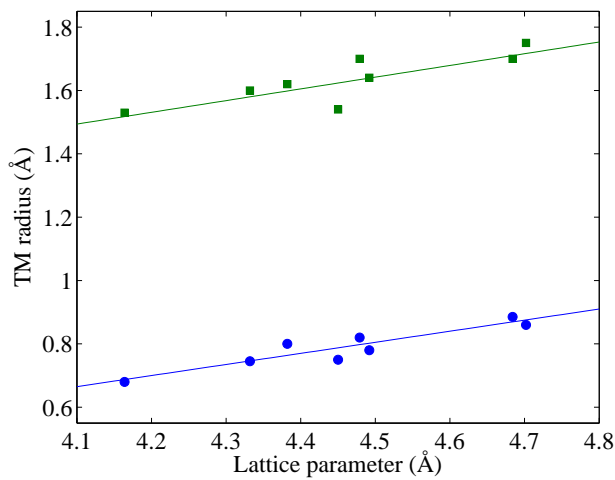}
\caption{\label{fig:a0_vs_rion_rcov}
  The linear correlations of the TMC lattice parameters with the
  covalent radii~\cite{CoGoPlReEscCrBaAl05} (squares) and with the
  ionic radii~\cite{Shannon76} (circles) of the parent transition
  metals.} 
\end{figure}

The trends in lattice parameters can be understood qualitatively by
plotting them against empirical covalent~\cite{CoGoPlReEscCrBaAl05}
and ionic~\cite{Shannon76} radii (Fig.~\ref{fig:a0_vs_rion_rcov}). The
overall linear correlations indicate that the bonding in the TMC's
contains both covalent and ionic effects. This iono-covalent mixture
agrees with earlier studies~\cite{Sch87} and is examined in more
detail in our analysis of the electronic structure below.   

The cohesive energies increase monotonically down a group in the
periodic table. Along each period, they exhibit a maximum for group
IV. These trends agree with the experimental values (see
Table~\ref{tab:TMC_a0}). As will be argued below, these trends can be
understood from the trends of the bulk electronic structure.  

%
%

\subsection{Valence Electron Density and Charge Transfer}\label{sec:dens_diff_Bader} 

%
\begin{figure*}
\centering
\includegraphics[width=0.43\textheight]{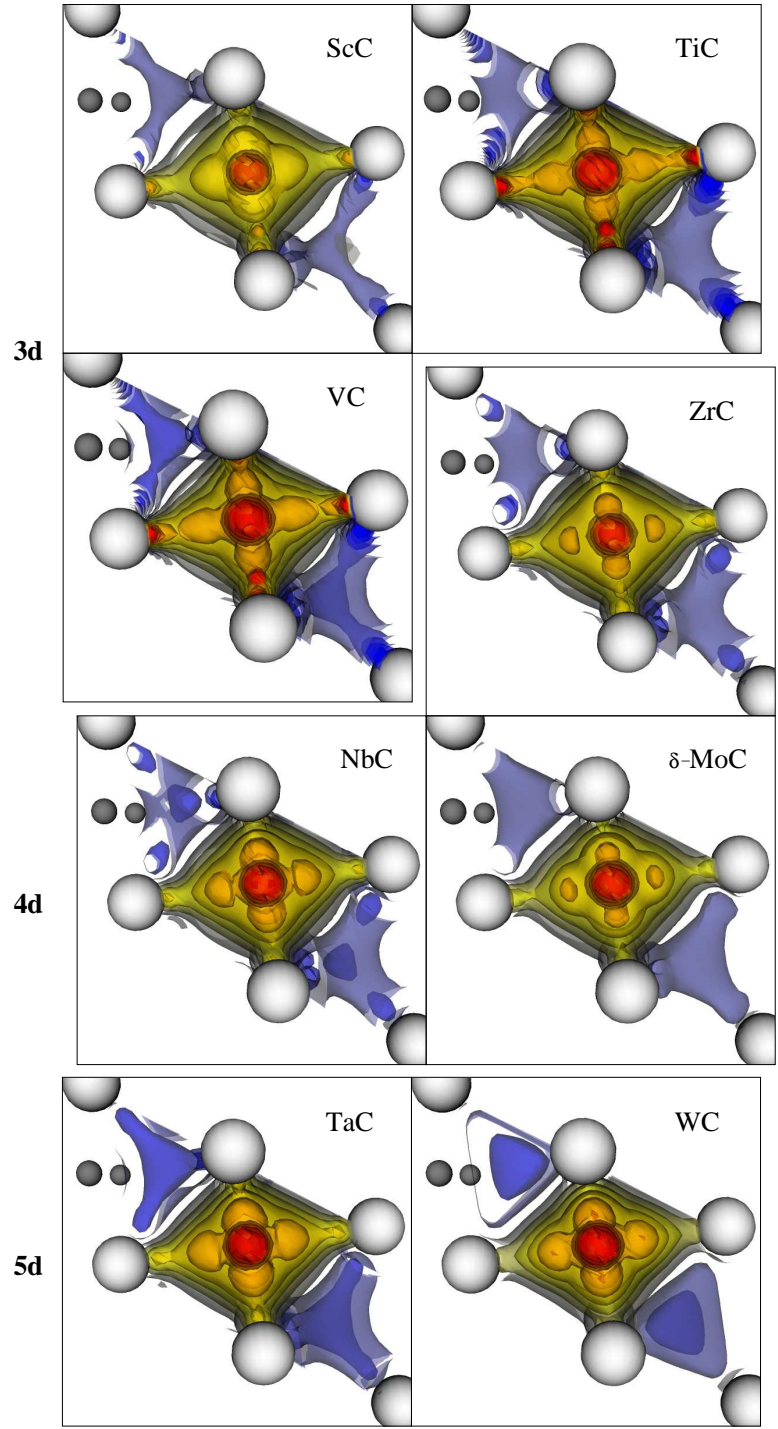}
\caption{\label{fig:dens_diff}
Valence-electron density difference plots for the considered bulk
systems, compared to the free constituent TM and C atoms. The
structures are viewed along the [111] direction. Twenty isosurfaces
homogeneously distributed in the interval  
$[-0.4,0.5]$ electrons/\AA$^3$ are shown. The color coding is blue
(reduction of electron density) through yellow (no density difference)
to red (increase in electron density). Larger (gray) balls correspond
to TM atoms and smaller black balls are C atoms. The atom situated in
the central red region is C.}   
\end{figure*}
%

Three-dimensional contour plots of the differences between the valence
electron densities in the bulk TMC's and the electron densities of
free TM and C atoms are shown in Fig.~\ref{fig:dens_diff} for the
considered TMC's. The changes in the distribution of the valence
electrons upon bond creation in the bulk TMC's reveal 
(i) a partially ionic character of the bond, seen as quite localized
electron clouds (orange-red regions) around the C atoms, indicating a 
charge transfer from TM to C, and (ii) covalent TM--C $\sigma$ bonds,
seen as a localization of charge (orange) in the region in between the
TM and C atoms. 

Our Bader analysis results presented in Table~\ref{tab:TMC_a0} confirm
that the TMC's are partly ionic compounds. Charge is transferred from
TM to C for all considered compounds, as expected from the higher
electronegativity of C. This direction of the charge transfer is
confirmed experimentally by the near-edge X-ray absorption fine
structure (NEXAFS) technique.\cite{Chen97}  

The covalent nature of the bonding is most evident in TiC and VC, as
shown by the charge-difference contours (Fig.~\ref{fig:dens_diff})
and by the lower Bader charge transfer for these compounds. In
general, the Bader charge trends indicate 
(i) an increase in covalency and a decrease in ionicity (an exception
is the metastable $\delta$-MoC), when moving from left to right along
a period, and (ii) a decrease in covalency and an increase in ionicity
down a group. 

All of the considered carbides except MoC and WC are stable in the
NaCl structure. The electron-density difference alone cannot explain
why this is the case. To gain an understanding of the structure change
of WC other methods must be employed, including the ones in the next
Section.

\subsection{Electronic Structure and Nature of Bonding}\label{sec:DOS_bulk} 

%
\begin{figure*}
\centering
\includegraphics[width=\textwidth]{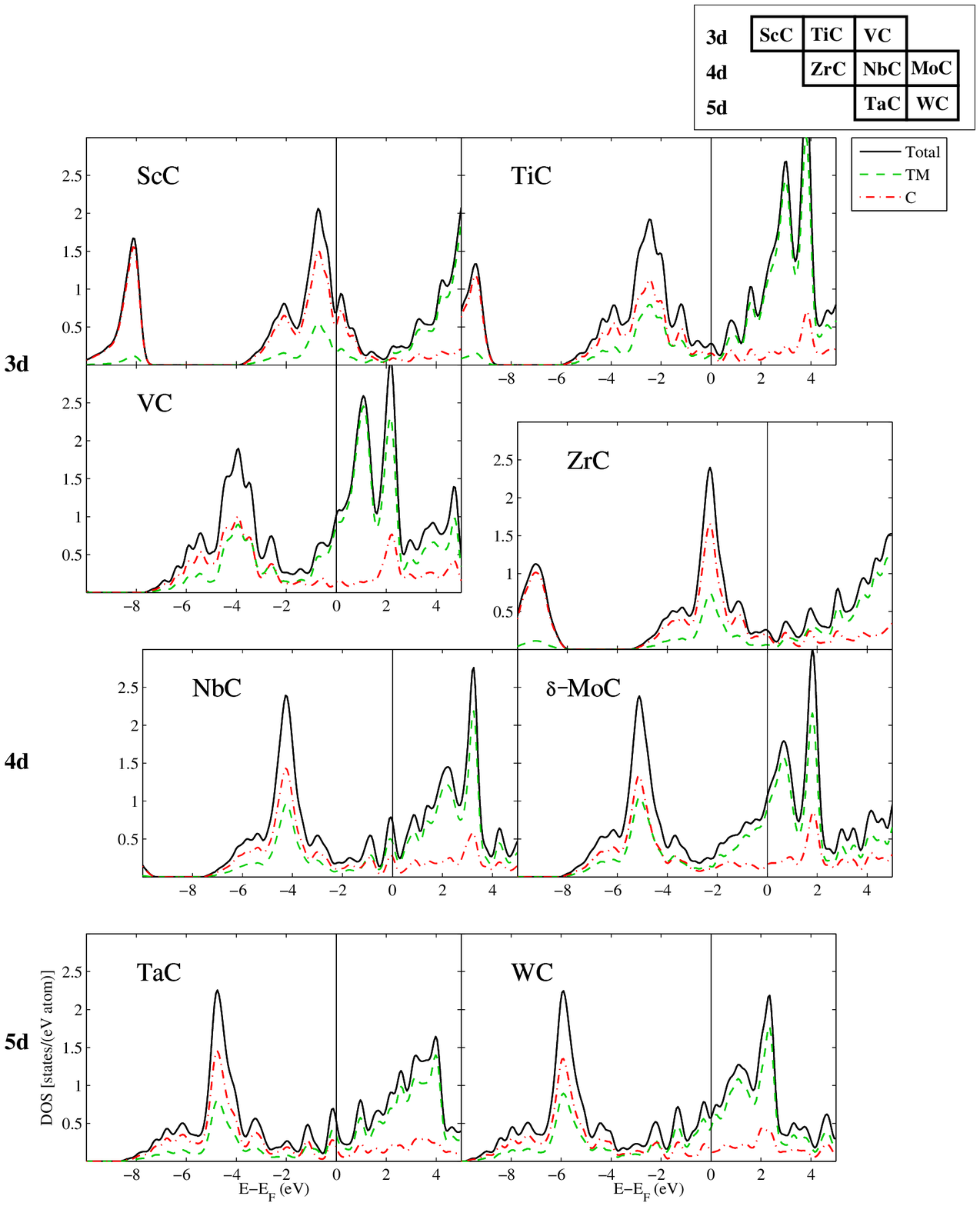}
\caption{\label{fig:DOS_bulk}
  Total and atom-projected densities of states for the considered bulk
  transition-metal carbide systems.} 
\end{figure*}
%

%
\begin{figure*}
\centering
\includegraphics[width=0.85\textwidth]{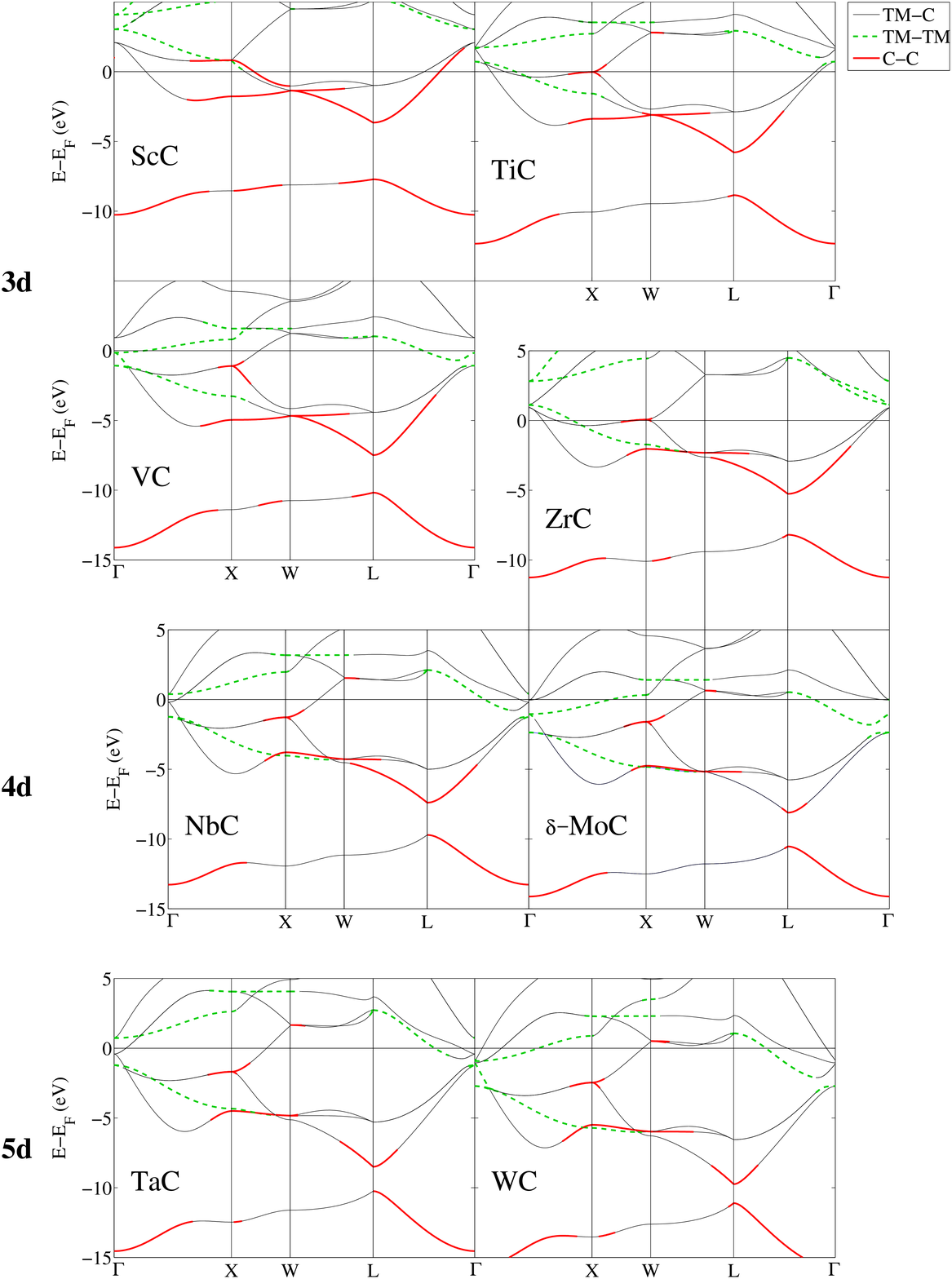}
\caption{\label{fig:band_structure}
  Calculated band structures for the considered TMC's. The parts of
  the bands that correspond to TM--TM, TM--C and C--C bonding states,
  respectively, are indicated. The identification of the bond
  character is based on the Kohn-Sham wave functions as described in
  the text and visualized in Fig.~\ref{fig:band_structure_psi}.}      
\end{figure*}
%

%
\begin{figure*}
\centering
\includegraphics[width=\textwidth]{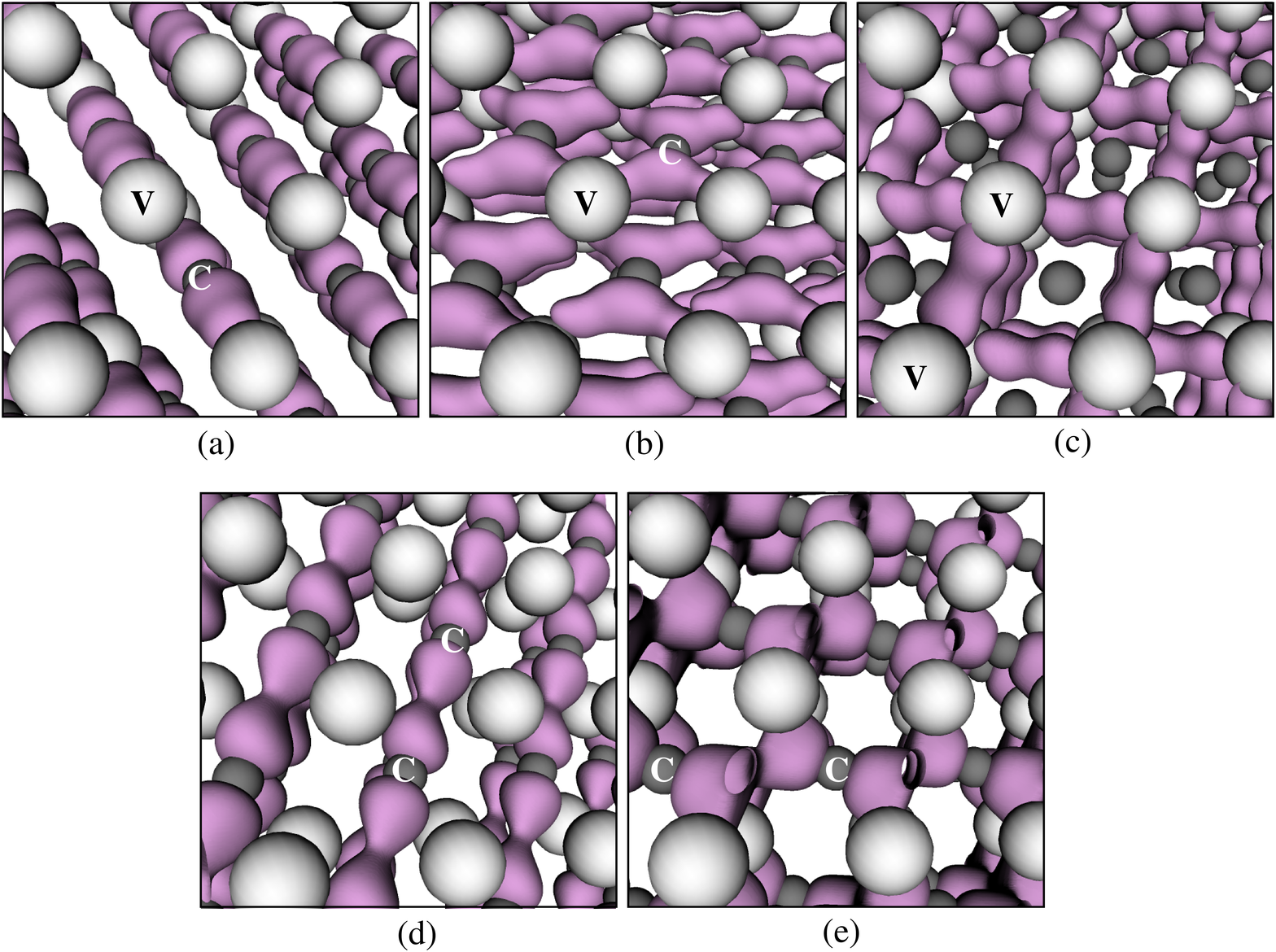}
\caption{\label{fig:band_structure_psi}
  Three-dimensional real-space contour plots of a representative
  selection of the Kohn-Sham wave functions 
  $|\Psi_{n\textbf{k}}(\textbf{r})|$ in bulk VC illustrating the different
  types of bonds at specific $\textbf{k}$ points and bands $n$ (here,
  $n=1$ corresponds to the lowest valence band,
  \textit{cf.}~Fig.~\ref{fig:band_structure}): 
  (a) TM($t_{2g}$)-C($p$) $\sigma$ bonds found in the lowest UVB band
  ($n=2$) between $\Gamma$ and $X$, 
  (b) TM($e_g$)-C($p$) $\pi$ bonds at $L_{n=3}$, 
  (c) TM($t_{2g}$)-TM($t_{2g}$) $\sigma$ at $X_{n=3}$, 
  (d) C($p$)-C($p$) $\sigma$ at $W_{n=2}$, (e) C($p$)-C($p$) $\pi$ at
  $L_{n=2}$ (there is no connection to the V atoms). The given contour
  plots are for the same value of
  $|\Psi_{n\textbf{k}}(\textbf{r})|=0.30$. Larger (gray) balls
  correspond to V atoms and smaller black balls ones to C atoms.}         
\end{figure*}
%

\subsubsection{General Features}

Calculated total and atom-projected DOS's and band structures
for the considered bulk TMC's are shown in Figs.~\ref{fig:DOS_bulk}
and \ref{fig:band_structure}, respectively. Furthermore, complementary 
information about the bond character for different bands and $k$ points 
is obtained by investigating the real-space Kohn-Sham wave functions
in detail. Such an analysis is performed for all the TMC's. The
results are presented in Fig.~\ref{fig:band_structure} and a
representative selection of the different bond types is illustrated in
Fig.~\ref{fig:band_structure_psi}.    

The overall electronic structure is similar for all considered TMC's
and is characterized by
(i) a deeply bound lower valence band (LVB) with C $2s$ states and a
very small amount of TM $d$ states (not shown in the DOS figures
for all the TMC's in the given energy interval);
(ii) a filled 
(or, for ScC, partly filled)
bonding upper valence band (UVB) below $E_F$, with
overlapping C $2p$ and TM $d$ states;
(iii) an empty or partly filled antibonding conduction band (CB)
above or around $E_F$, dominated by TM $d$ states with a clear
contribution from C $2p$ states; and
(iv) a non-vanishing DOS in the pseudogap between the UVB and the
CB.

The contribution to the bonding from the LVB is presumably small due
to the small hybridization of the TM and C states. Points (ii) and
(iii) illustrate the presence of a covalent bond. The mainly C $2p$
and TM $d$ characters of the UVB and CB, respectively, indicate a
partially ionic bonding nature. 
Orbital-projected DOS's show that the TM states are of predominantly 
$e_g$ and $t_{2g}$ symmetry in the UVB and in the CB, respectively. 
Property (iv) explains the metallic character of the TMC's. 

An analysis of the Kohn-Sham wave functions along the $k$ symmetry
lines of Fig.~\ref{fig:band_structure} shows that the band structure
is dominated by TM--C states [see
Figs.~\ref{fig:band_structure_psi}(a)-(b) for representative
TM--C bonds in the case of VC]. These states are found throughout
the whole energy region of the $k$ space, that is, they are
present in both the LVB and the UVB, as well as in the CB.

The wave function analysis reveals also that large parts of
the states in the LVB and in the lower energy range of the UVB have a
bonding C--C character of both $\sigma$ and $\pi$ symmetry
[see Figs.~\ref{fig:band_structure_psi}(d)--(e) for representative
C--C bonds in the case of bulk VC].

Finally, TM--TM states are evident in significant parts of both
the UVB and the CB
[see Fig.~\ref{fig:band_structure_psi}(c) for a representative TM--TM
bond in the VC system].  In all the TMC's, at least one band of
TM--TM character crosses $E_F$.

The wave function analysis confirms also the picture that the UVB
states have a bonding symmetry, while the CB states show a more
antibonding character (\textit{i.e.}, nodes in between the atoms). 

Experiments on the band structures for single bulk crystals of group
IV-VI TMC's have been carried out by using, for example,
angle-resolved photo-emission spectroscopy (ARPES), in particular
X-ray photoemission spectroscopy (XPS) for the mapping of occupied
bands and inverse photo-emission spectroscopy (IPES) for the mapping
of unoccupied bands (see Ref.~\onlinecite{Jo95} and references
therein). These results agree with the rough features of our
calculated results but do not provide any information on the different
bond characters of the different bands, which our method is able to
give.

\subsubsection{Trends along a period}
The DOS figures, together with the band structure and wave
function analyses, show that, as the TM group number increases,
(i) both the UVB and the CB (as well as the LVB)
are shifted down in energy relative to $E_F$, due to the filling
of bands;
(ii) as a consequence the metallicity (\textit{i.e.}, the number of
states at $E_F$) increases;
(iii) the bands crossing $E_F$ 
show more TM--TM character;
(iv) the energy separation between the UVB and the CB decreases;
(v) the UVB becomes less C localized; and
(vi) the overall amount of C--C states decreases.

\subsubsection{Trends along a group}
As the TM period number increases the following general features are
observed in the DOS, band structure, and wave function plots:
(i) the position of $E_F$ relative to the center of mass of the UVB
and the CB is unchanged;
(ii) the bond character of the different states is largely unchanged;
(iii) the energy separation between the UVB and the CB increases; and
(iv) the UVB becomes more C localized. The differences between periods
5 and 6 are smaller than the differences between periods 4 and 5.

Also, a detailed investigation of the band structure reveals that the
energetical order of the bands changes: 
(i) at the $\Gamma$ point the higher TM--TM band moves up in energy,
while TM--C bands move down; and 
(ii) the two lowest UVB bands at the $X$ point (the C--C and the
TM--TM bands) approach each other and change places.

\subsection{Understanding the trends}

%
\begin{figure}
\includegraphics[width=.4\textwidth]{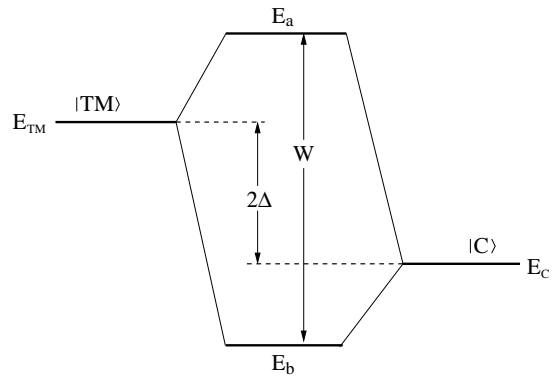}
\caption{An illustration of a two-level system consisting of the
  valence TM and C states. The bonding $E_{\text{b}}$ and antibonding
  $E_{\text{a}}$ states correspond to the UVB and CB of the TMC,
  respectively.}  
\label{fig:twolevel}
\end{figure}
%

The above trends can be explained on the basis of a two-level system,
comprised of the valence $d$ and $p$ states of the free TM and C
atoms, respectively (see Fig.~\ref{fig:twolevel}). This discussion
resembles the one for a heteronuclear diatomic molecule (see for
example Ref.~\onlinecite{Subook}). Due to the lower electronegativity
of the TM atom, the energy of the TM $d$ state is higher than that of
the C $p$ state. Their interaction results in bonding $E_{\text{b}}$
and antibonding $E_{\text{a}}$ states, the UVB and the CB,
respectively, separated by an energy $W$ (see
Fig.~\ref{fig:twolevel}). Assuming a one-electron picture, the
solutions to the Schr\"{o}dinger equation are  
\begin{align}
&E_{\text{b}}=\varepsilon-\sqrt{\Delta^2+\beta^2}\\
&E_{\text{a}}=\varepsilon+\sqrt{\Delta^2+\beta^2},    
\end{align}
with $\varepsilon=(E_{\text{TM}}+E_{\text{C}})/2$ and
$\Delta=(E_{\text{TM}}-E_{\text{C}})/2$, where $E_{\text{TM}}$ is the
energy of the TM $d$ level and $E_{\text{C}}$ is the energy of the C
$p$ level, and $\beta=\langle C|H_{12}|TM\rangle$. If $\Delta\neq0$,
which is the case here, there is a charge transfer from TM to C and
the bond is partially ionic.  

As the TM group number increases, the TM level is shifted down in
energy, due to the increase of $d$ electrons, giving a better energy
overlap between the TM and C states. This results in a stronger $pd$
hybridization, a smaller $\Delta$, and a bond of more covalent
character, that is, a more similar amount of TM and C localization of
the UVB. At the same time, the increase of $d$ electrons causes a
shift of $E_F$ towards higher energies. The separation between UVB and
CB is given by $W=2(\Delta^2+\beta^2)^{1/2}$. Along a period, each
extra $d$ electron is added in the same valence shell. To a first
approximation one can therefore assume that $\beta$ is constant along
a period. Hence, $W$ becomes smaller, which is confirmed by our DFT
calculations (see Fig.~\ref{fig:DOS_bulk}).  

Moving down a group, the number of $d$ electrons is constant, leaving
$E_F$ at the same energy. However, the TM level is shifted up in 
energy, resulting in a smaller $pd$ hybridization and a larger
separation between the TM and the C levels, giving a more ionic
bond. As the separation increases, the closer the resulting levels
come to the original ones, giving them more character of the original
levels, that is, the UVB (CB) becomes more C (TM)-localized. 

This picture is supported by our calculated Bader charge-transfer
trends (Table~\ref{tab:TMC_a0}). To the right along a period, the
ionicity decreases, which results in decreasing Bader values (except
for $\delta$-MoC, as mentioned in
Sec.~\ref{sec:dens_diff_Bader}). Down a group, the increase in
ionicity is reflected by increasing Bader values.  



\section{Discussion}\label{sec:disc}
In this paper we present a systematic and detailed trend study of
the electronic structure of a collection of early transition-metal
carbides based on DFT calculations.

The results confirm the mixed iono-covalent TM--C and the metallic
TM--TM nature of the bond. This is shown both by local DOS analyses
and by a thorough analysis of the band structures and Kohn-Sham wave
functions along several symmetry lines in the Brillouin zone
(Fig.~\ref{fig:band_structure}). TM--C bands are present throughout
the whole energy range of the \textit{k} space, including the
pseudogap between the UVB and the CB. Also, both a bonding and an
antibonding TM--TM band are identified. 

In addition, our analysis shows that in all the considered TMC's there
exist direct C--C bonds, localized mainly in the lower part of the
UVB. This presence is most pronounced in ScC (where C--C bands cross
$E_F$) and decreases as the TM group and period numbers increase. The
C--C bonds are also found in the pseudogap. 

Hence, the C atoms play two roles for the bonding in the TMC's, 
by their interaction with the TM atoms and with other C atoms,
respectively. 

The valence-electron density difference analysis shows that the
dominating bond character is the iono-covalent TM--C bond, which is
supported by the mapping of bands in the band structure (see
Fig.~\ref{fig:band_structure}). The trends in the covalent and ionic
contributions can be extracted even though these two bonding types are
closely intertwined. We find that the covalent contribution increases
from left to right along a period and decreases down a group, while
the ionic one decreases from left to right and increases down a group.
In addition, our results show that there is a correlation between the
covalency and the separation between the UVB and CB. A larger
covalency corresponds to a larger separation. 

Although the TM--TM and C--C bonds are not observed in the density
difference plots, their existence is revealed by the real-space
wave function analysis. A contribution from C--C bonds to the TMC
cohesion has been suggested in Ref.~\onlinecite{ZhLiZhXi02} based on a
two-sublattice model. However, such a model does not take into account
the modifying effect of the TM $d$ states on the C orbitals in the C
sublattice, and vice versa. Our approach identifies the different
bonding types directly from the calculated electronic structure and
therefore includes this modifying effect. We find that a significant
amount of C--C states are present in all the studied TMC's, that is,
not only the metastable ones, as was suggested in
Ref.~\onlinecite{ZhLiZhXi02}.   

All the studied TMC's have a metallic bond character, that is, a
non-vanishing DOS at $E_F$. These states
(Fig.~\ref{fig:band_structure}) are of both TM--TM and TM--C character
(for ScC also C--C states). Indeed, measurements show 
that group IV and V TMC's are almost as good electrical conductors as
the parent metals.\cite{Pi96} With an increasing DOS at $E_F$ 
(Fig.~\ref{fig:DOS_bulk}), and assuming everything else weakly
varying, the conductivity should increase from group IV to group
VI, which agrees with measurements.\cite{Pi96} On the same arguments,
the conductivity of ScC should be comparable to that of VC.

As discussed in Sec.~\ref{sec:bulk_geometry}, along a period the
calculated cohesive energies exhibit a maximum for group IV. The
calculated electronic structure shows that for group IV, $E_F$ is
positioned in the pseudogap between the UVB and the CB, meaning that
all the bonding UVB states are filled and all the antibonding CB
states are empty. The same arguments as in the case of the bonding
energy in a diatomic molecule can be applied, that is, a filling of
bonding states increases the bond strength, while filling of  
antibonding states reduces the bond
strength.\cite{NeRaEiWeSch76,GeWiMo83,Sch87,ZhGuJeChAn88,HaGuGrKo93} 
For ScC, not all bonding UVB states are filled and therefore the
cohesive energy is lower for this compound. For groups V and VI, the
antibonding CB states are partly filled, which again results in a
weaker bond.   

Down a group, the cohesive energies of the TMC's increase. This is
because the UVB is shifted down in energy, making the bond stronger,
while the CB is shifted up in energy, resulting in an energy gain due
to the emptying of antibonding states.    

The band structure and Kohn-Sham wave function analyses show the
origin of the instabillity of MoC and WC in NaCl structure.
The instability arises from the filling of antibonding states, as
shown in Fig.~\ref{fig:band_structure}. In MoC, the TM--TM
antibonding band around $\Gamma$ is partly filled. In WC, two
additional antibonding bands, of W--C character, cross $E_F$ around
the $\Gamma$ point and become partly occupied. This picture agrees
with previous studies~\cite{PrCo89} but also provides additional
details.  

The detailed analysis of the electronic structure provided in this
study lays the essential foundation for the understanding of the
TMC's surface properties, such as their surface reactivity. The surface  
characteristics of a material are tightly bound to its underlying bulk
properties, and will of course depend strongly on the type of
bonds that are broken upon creation of the surface. It is, for
example, known that the surface electronic structure of the TMC (100)
surfaces is similar to that of the bulk. However, the TMC (111)
surface shows both TM-localized as well as C-localized surface
resonances.\cite{RuLu07,VoHeRuLu,VoRuLu} The importance of these
resonances on the surface reactivity is investigated in
Refs.~\onlinecite{VoHeRuLu} and~\onlinecite{VoRuLu}.             



\section{Conclusions}\label{sec:conc}

This trend study deals with the bonding nature of early TMC's. 
Our approach is based on a complementary use of different types of
electronic-structure analysis tools. In particular, a thorough mapping
of the band structure provides detailed insight into the bonding of
the bulk TMC's. The results
(i) confirm that the dominant contribution to the bond is the
iono-covalent TM--C bond, 
(ii) show the existence of TM--TM bonds, and, importantly, 
(iii) reveal the existence of direct C--C bonds (most pronounced for
ScC). We provide new information on the spatial extent of the
different bonds, on their \textit{k}-space location within the band 
structure, and on their importance for the bulk cohesion. Also, trends
in covalency \textit{vs.}~ionicity are obtained.
The resulting electron-structural trends are analyzed and discussed
within a two-level model. 

These results are of importance for the understanding of the TMC
surface reactivities, due to the intimate relation between bulk and
surface electronic structures. When creating a surface by cutting a
crystal, the breaking of TM--C, TM--TM, and C--C bonds can manifest
themselves as surface resonances and/or surface states at the
surface. As we show in another study, not only TM but also C
resonances play a crucial role in the bonding mechanism on TMC
surfaces.\cite{VoHeRuLu,VoRuLu}



\begin{acknowledgments}

The authors thank B.~I. Lundqvist and A. Hellman for reading the
manuscript and for providing valuable suggestions. Allocation of
computer time via SNIC (Swedish National Infrastructure for Computing)
is gratefully acknowledged.    

\end{acknowledgments}




\end{document}